\documentclass[10pt,a4paper]{iopart}
\usepackage{amstext}

\usepackage[varg]{txfonts}   
\usepackage{textcomp}
\usepackage{graphicx}

\usepackage[numbers]{natbib}

\usepackage[colorlinks=false,
            bookmarks=true,
            bookmarksnumbered=true]{hyperref}
\hypersetup{pdfborder = {0 0 0}}

\csname endofdump\endcsname

\renewcommand{\etal}{\textit{et al}}
\newcommand{\aSnSe}{\mbox{$\alpha$-SnSe}}
\newcommand{\bSnSe}{\mbox{$\beta$-SnSe}}
\renewcommand{\vec}[1]{\ensuremath{\mathbf{#1}}}%

\begin{document}

\title{Structural Changes in Thermoelectric SnSe at High Pressures\hspace*{-10mm}}

\author{I Loa$^1$, R~J~Husband$^1$, R~A~Downie$^2$, S~R~Popuri$^2$, J-W~G~Bos$^2$ }

\address{$^1$SUPA, School of Physics and Astronomy, and Centre for Science
             at Extreme Conditions, The University of Edinburgh, Edinburgh, EH9 3FD, United Kingdom\\
         $^2$Institute of Chemical Sciences and Centre for Advanced Energy Storage and Recovery,
         School of Engineering and Physical Sciences, Heriot-Watt University, Edinburgh, EH14 4AS, United Kingdom}
            \ead{I.Loa@ed.ac.uk}

\vspace{10pt}
\begin{indented}
\item[] \today
\end{indented}

\begin{abstract}
The crystal structure of the thermoelectric material tin selenide has been investigated with
angle-dispersive synchrotron x-ray powder diffraction under hydrostatic pressure up to 27~GPa.
With increasing pressure, a continuous evolution of the crystal structure from the GeS type to
the higher-symmetry TlI type was observed, with a critical pressure of 10.5(3)~GPa. The
orthorhombic high-pressure modification, $\beta{'}$-SnSe, is closely related to the
pseudo-tetragonal high-temperature modification at ambient pressure. The similarity between
the changes of the crystal structure at elevated temperatures and at high pressures suggests
the possibility that strained thin films of SnSe may provide a route to overcoming the problem
of the limited thermal stability of $\beta$-SnSe at high temperatures.

\end{abstract}

%
%
%
%

%

\section{Introduction}

The central challenge in finding or engineering thermoelectric materials suitable for thermal
energy conversion beyond niche applications remains to combine good electrical properties and
a very low thermal conductivity in a single material. Very recently, the compound SnSe has
been reported to exhibit exceptionally good thermoelectric properties at high temperatures
above $\sim$800~K, including a very low thermal conductivity \cite{ZLZS14}. Nanostructuring of
materials has been pursued extensively in recent years as a possible route towards achieving a
sufficiently low thermal conductivity \cite{ST08,VP10,BHBW12}, but the discovery of extremely
low thermal conductivity in bulk SnSe has markedly improved the prospects of obtaining bulk
thermoelectrics suitable for general commercial applications. As SnSe consists only of
Earth-abundant elements of comparatively low toxicity, it represents also an important advance
in avoiding toxic and rare elements like lead and tellurium in thermoelectric materials.

\begin{figure}
	\centering
	\includegraphics[scale=1.0]{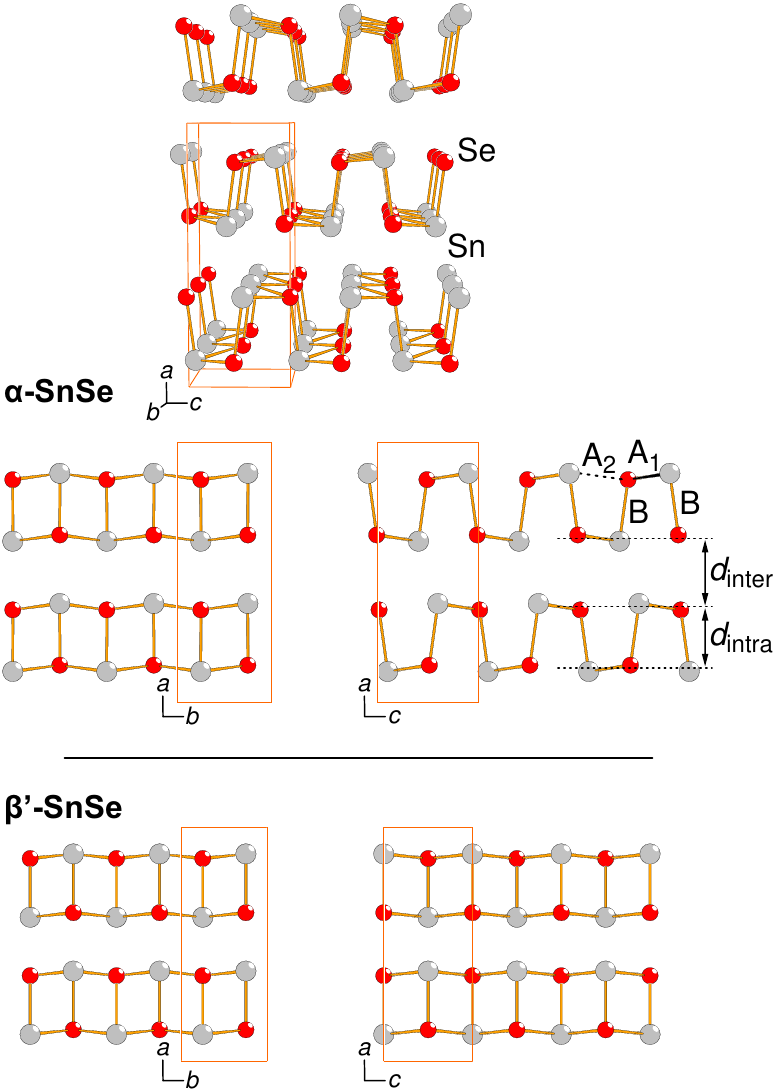}\\
	\caption{Crystal structures of $\alpha$-SnSe at 0 GPa (top)  and $\beta{'}$-SnSe at
27~GPa (bottom). The light rectangles indicate the unit cell.}
	\label{fig:crystal_structure}
\end{figure}

At ambient conditions, SnSe adopts a layered orthorhombic crystal structure (GeS structure
type; space group $Pnma$, No.\ 62) as illustrated in Fig.~\ref{fig:crystal_structure}. The
structure can be regarded as a heavily-distorted variant of the NaCl structure type, and it
comprises NaCl-type double layers that are stacked along the $a$ direction. One signature of
the distortion are the unequal Se--Sn interatomic distances A1 and A2, which gives rise to
{``}corrugated{''} layers as seen in projection along the $b$ direction, see
Fig.~\ref{fig:crystal_structure}. Another is that adjacent NaCl-type double layers are
displaced by approximately $\vec{c}/2$ with respect to each other. With increasing
temperature, the structure was observed to evolve continuously into a higher-symmetry variant
(TlI structure type; space group $Cmcm$, No.\ 63), with a critical temperature of 807~K
\cite{WC79,SW81}. It is this high-temperature, higher-symmetry phase of SnSe that has the
excellent thermoelectric properties. Unfortunately, SnSe appears to be thermally rather
unstable under these conditions due to sublimation \cite{SW81}, which severely limits the use
of this material in devices.

Here, we investigate the effect of high hydrostatic pressure on the crystal structure of SnSe.
Our study was motivated (i) by recent reports that some thermoelectric materials exhibit
dramatic improvements in the thermoelectric power factor under compression \cite{OS07,OSVM08},
(ii) by the fact that earlier studies (as detailed below) have yielded only a rather
fragmentary and partially contradictory picture of the behaviour of SnSe under high pressure
and (iii) by the scarcity of state-of-the-art experimental crystal structure investigations on
group-IV chalcogenides in relation to the abundance of computational studies. Using
synchrotron powder x-ray diffraction experiments, we quantify the structural changes in SnSe
under pressure and demonstrate that there is a pressure-induced transition to the
higher-symmetry GeS structure type, and we comment on the possible implications of the
structural changes for the thermoelectric properties of SnSe.

\section{Earlier High-Pressure Studies of SnSe}

 The properties of SnSe under high pressure have been the subject of a number of
experimental and computational studies, largely motivated by SnSe and related orthorhombic
IV-VI compounds being intermediate between two- and three-dimensional solids in terms of their
anisotropic physical properties. Chattopadhyay \etal\ \cite{CWSP84} investigated the crystal
structure of SnSe under high pressure to 34~GPa by energy-dispersive x-ray diffraction.
Judging by the absence of discontinuities in the interplanar distances, they found no evidence
for a structural phase transition. However, they noted the limitations of the energy
dispersive \mbox{x-ray} diffraction technique, suggesting that a continuous and relatively
subtle transition such as the $Pnma \to Cmcm$ transition known from high-temperature
experiments might be difficult to detect with this technique.

In a high-pressure M\"{o}ssbauer study \cite{PM90}, Peters and McNeil identified the low-pressure
range below 3.0~GPa as the range in which the interlayer bonds are changed, transforming SnSe
from a predominantly two-dimensional to a more three-dimensional material. Parenteau and
Carlone \cite{PC90} measured the temperature and pressure dependence of the direct and
indirect optical gaps of SnSe up to 4.0~GPa. They found its electronic properties to be
three-dimensional in nature, already at ambient pressure, as opposed to the two-dimensional
nature suggested by the easy cleavage of SnSe crystals, their lattice dynamics \cite{CHZC77}
and the interpretation of the chemical bonding in this compound \cite{WC79}. Agarwal \etal\
\cite{ACPL94,ATL05} observed abrupt changes in the electrical resistivity and thermopower near
6~GPa, but their exact origin has remained unclear. Interestingly, the resistivity decreases
monotonically with increasing pressure, whereas the thermopower decreases up to 6.3~GPa,
followed by large enhancement between 6.5 and 7.6~GPa.

Two recent computational studies also investigated the structural and electronic properties of
SnSe under pressure. Alptekin \cite{Alp11} predicted a pressure-induced second-order
transition to a phase with space group $Cmcm$, isostructural with the high-temperature phase
\bSnSe. Two possible phase transition pressures of 2 or 7 GPa were identified, depending on
the method of calculation. In this study, little evidence of pressure-induced electronic
changes was found; in particular, SnSe was reported to remain semiconducting in the $Cmcm$
phase at 7~GPa. In contrast, Makinistian and Albanesi \cite{MA11} reported a reduction of the
band gap with increasing pressure and a transition to a semimetallic state with indirect band
overlap. As the calculations underestimate the zero-pressure band gap by $\sim$50\% in
comparison with experimental results, the calculated metallisation pressure of $\sim$5~GPa
should be regarded as a lower bound.

\section{Experimental Details}

Polycrystalline SnSe was synthesised using a standard solid state chemistry reaction.
Stoichiometric amounts of 100-mesh Sn powder (99.9\% stated purity) and shots of Se (99.999\%)
were ground together and cold-pressed into a pellet. The pellet was sealed in an evacuated
quartz tube and heated at 900~{\mbox{\textdegree}}C for 24 hours (10~{\mbox{\textdegree}}C/min heating ramp), followed by cooling at
3~{\mbox{\textdegree}}C/min to 600~{\mbox{\textdegree}}C, after which the oven was switched off. This heating step was repeated once
with an intermediate homogenisation. Phase purity was checked using a Bruker D8 Advance
diffractometer with monochromated Cu-K$_{\alpha1}$ radiation.

Post-synthesis hot-pressing using a home-built apparatus (500\;{\mbox{\textdegree}}C and 80~MPa for 20~minutes)
was used to obtain a fully dense SnSe sample. The electrical resistivity and Seebeck
coefficient were measured using a Linseis LSR3 instrument and found to be 5~$\Omega$\,cm and
$+490$~{\mbox{\textmu}}V\,K$^{-1}$ at 300~K. The Seebeck coefficient agrees with recently reported values for
spark-plasma-sintered polycrystalline SnSe \cite{SCVO14}, whereas our resistivity is a factor
of 10 larger.

The synthesised material was manually ground to a fine powder and loaded into a
Merrill-Bassett-type diamond anvil cell (DAC) for high-pressure generation. Condensed helium
was used as the pressure-transmitting medium, and pressures were determined with the ruby
fluorescence method using the calibration of Ref.~\cite{MXB86}. Monochromatic powder x-ray
diffraction patterns were measured in the Debye-Scherrer geometry on beamline ID09a of the
European Synchrotron Radiation Facility (ESRF). Monochromatic x-rays of wavelength $\lambda =
0.4154$~{\AA} were focused to a spot size of $\sim$20~{\mbox{\textmu}}m at the sample, and the two-dimensional
diffraction images were recorded with a mar555 area detector. The DAC was rotated by ${\pm}3${\mbox{\textdegree}}
during the exposure to improve the powder averaging. The diffraction images were integrated
azimuthally with the Fit2D software \cite{soft:Fit2d} to yield conventional intensity vs
$2\Theta$ diffraction diagrams, which were analysed with the Rietveld method using the program
GSAS \cite{soft:GSAS,soft:EXPGUI}.

\section{Results and Discussion}
\subsection{Pressure-Induced Phase Transition}
\begin{figure}[bt]
     \centering
	  \includegraphics[scale=1.0]{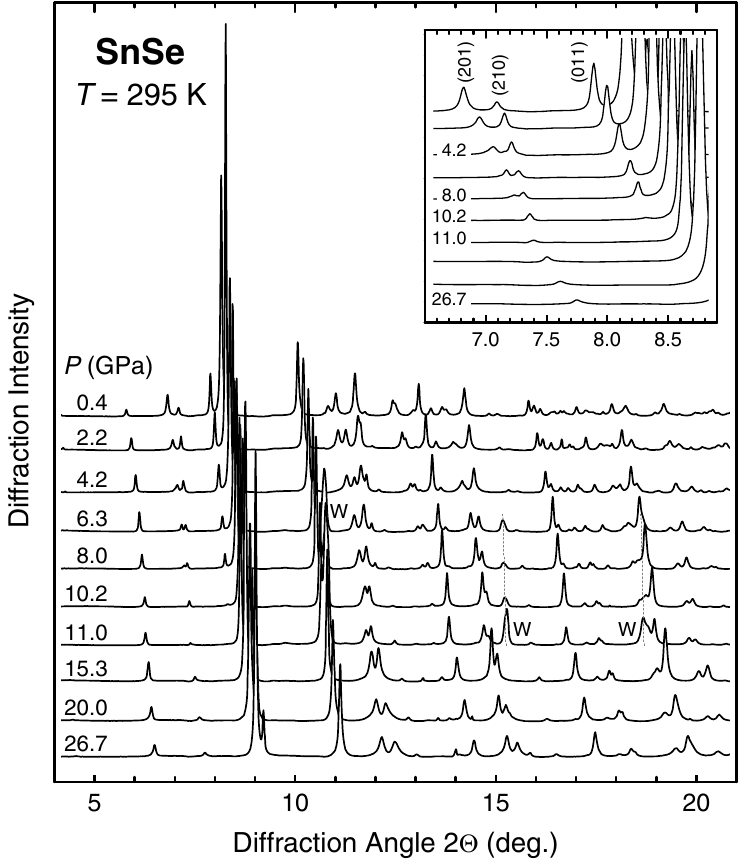}\\
     \caption{Selected powder x-ray diffraction patterns of SnSe at high pressures up to
              26.7~GPa. Diffraction peaks marked {`}W{'} near 10.7, 15.2 and 18.7{\mbox{\textdegree}} are from
              the tungsten gasket (due to a slight misalignment of the sample that
              remained unnoticed for several pressure steps). A smooth background due to
              Compton scattering from the detector-facing diamond anvil has been
              subtracted. The inset
             illustrates the pressure-induced reduction in intensity and
              disappearance of the (011) and (201) reflections.}
     \label{fig:xray_patterns}
\end{figure}

Figure~\ref{fig:xray_patterns} illustrated the evolution of the diffraction pattern of SnSe
for increasing pressures up to 27~GPa. No abrupt changes were observed, but there is a gradual
disappearance of several reflections as shown in the inset of Fig.~\ref{fig:xray_patterns} for
the (011) and (201) reflections. This suggest a transition from the structure with space group
$Pnma$ to a higher-symmetry variant, and this transformation is completed at 11.0 GPa.

\begin{figure}[bt]
     \centering
	  \includegraphics[scale=1.0]{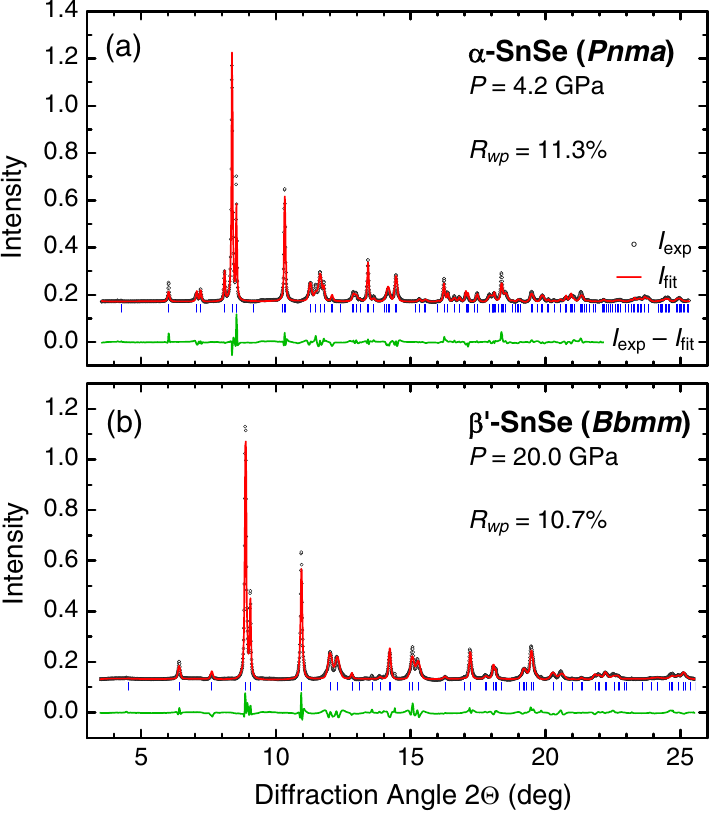}\\
     \caption{Rietveld fits for (a) the low-pressure phase $\alpha$-SnSe at 4.2~GPa and
              (b) the high-pressure phase $\beta{'}$-SnSe at 20.0~GPa. The tickmarks
              indicate the calculated peak positions.}
     \label{fig:Rietveld}
\end{figure}

The powder diffraction patterns were analysed with the Rietveld method. Starting at low
pressure, the crystal structure was first analysed based on the GeS-type structure model with
space group $Pnma$. The refined parameters were the lattice parameters, the $x$ and $z$ atomic
coordinates of Se and Sn, profile parameters of the Stephens peak function \cite{Ste99} and
the coefficients of a Chebyshev polynomial for the background. As the diffraction images
showed textured powder rings, a correction for preferred orientation of the powder particles,
based on spherical harmonics \cite{Dre97}, was applied. Even a conservative two-parameter
correction for preferred orientation improved the agreement between measured and fitted
intensities noticeably. All Rietveld refinements were performed with spherical-harmonics
corrections of orders 2, 4, and 6 (2, 5, 9 refinable parameters, respectively). The resulting
atomic coordinates from the three fits generally agreed within the estimated uncertainties,
and the results reported here are the averages of the three refinements for each diffraction
pattern. Figure~\ref{fig:Rietveld}(a) shows the Rietveld fit for $\alpha$-SnSe at 4.2~GPa as
an example. A spherical-harmonics correction of order 2 was used in this case, and the
weighted residual of the fit without background was $R_{wp}=11.3\%$.

\begin{figure}[bt]
     \centering
	  \includegraphics[scale=1.0]{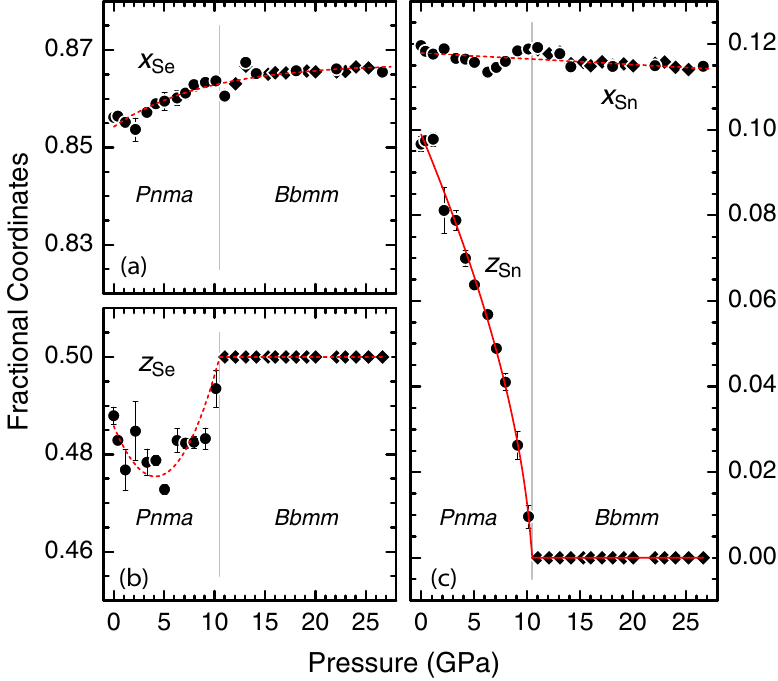}\\
     \caption{(a--c) Fractional atomic coordinates in SnSe as a function of pressure.
              Circles and diamond symbols show the results of Rietveld refinements based
              on the $Pnma$ and $Bbmm$ structures, respectively. Dashed lines are guides
              to the eye; the solid line for the Sn $z$-coordinate represents a power-law
              fit as described in the text. The grey vertical line marks the critical
              pressure of 10.5~GPa.}
     \label{fig:atom_coords}
\end{figure}

Figure~\ref{fig:atom_coords} shows the variation of the atomic coordinates with pressure as
obtained from the Rietveld refinements. Most notable is the clear evolution of the Sn
$z$-coordinate towards zero, a process which is again completed at 11~GPa. Attempts to refine
the structures for pressure above 11 GPa based on the $Pnma$ structure model led to rather
unstable fits, but the results for the Sn $z$-coordinate were consistent with $z=0$. The
results for the pressure dependence of the Se $z$-coordinate are less conclusive because it is
close to the high-symmetry value of $1/2$ already at ambient pressure, and the lower
scattering power of Se compared to Sn leads to larger uncertainties in the measured parameters
for the former. The data in Fig.~\ref{fig:atom_coords}(b) suggest that the Se $z$-coordinate
approaches 1/2 at the critical pressure, possibly after passing through a minimum near 5~GPa.
In addition, the absence of any traces of the (011) and (201) reflections at pressures above
11 GPa indicates that the $z$-coordinates of both Sn and Se adopt the high-symmetry values of
0 and $1/2$, respectively. This changes the space group of the crystal structure from $Pnma$
(\#62) to $Bbmm$ (\#63). The latter is a non-standard setting of space group $Cmcm$ that keeps
the crystal axes in the same orientation as for $Pnma$. This pressure-induced change in
symmetry is the same as that observed upon heating at ambient pressure. For reasons explained
below along with a discussion of the relation between the temperature- and pressure-induced
structural changes, we denote the high-pressure phase as $\beta{'}$-SnSe.

For the data collected at pressures of 11~GPa and higher, Rietveld refinements of the crystal
structure were performed with space group $Bbmm$. Figure~\ref{fig:Rietveld}(b) shows the
Rietveld fit for $\beta{'}$-SnSe at 20.0~GPa as an example. As before, a preferred-orientation
correction of order 2 was applied, and the weighted residual of the fit without background was
$R_{wp}=10.7\%$.

The atomic coordinates obtained from the $Bbmm$-based refinements are shown with
diamond-shaped symbols in Fig.~\ref{fig:atom_coords}. Within the experimental uncertainty, the
$x$-coordinates of Se and Sn change continuously over the whole pressure range and thus across
the transition near 11~GPa. The pressure dependence of the Sn $z$-coordinate below the
transition pressure can be described well with a power law,
\begin{equation}
   z_\text{Sn}(P) = z_\text{Sn}(0) \left( \frac{P_c - P}{P_c}\right)^\alpha,
\end{equation}
with $z_\text{Sn}(0) = 0.0996(13)$, a critical pressure $P_c = 10.5(3)$~GPa and a critical
exponent $\alpha=0.64(4)$.

\subsection{Anisotropic Compressibility}

\begin{figure*}[t]
     \centering
	  \includegraphics[scale=1.0]{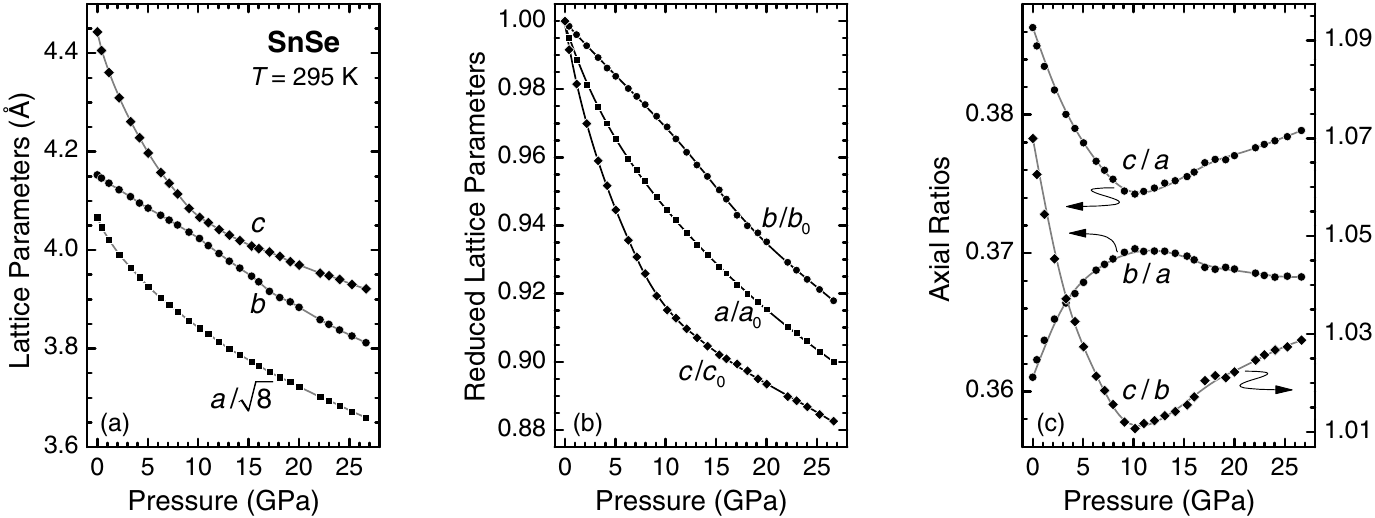}\\
     \caption{(a) Lattice parameters and (b) reduced lattice parameters of SnSe as a
              function of pressure. The zero-pressure lattice parameters are
              $a_0=11.5022(15)$~{\AA}, $b_0 = 4.1526(9)$~{\AA} and $c_0=4.4432(7)$~{\AA}. (c) Axial
              Ratios $b/a$, $c/a$ and $c/b$ as a function of pressure. The symbols show
              the experimental data and lines are guides to the eye.}
     \label{fig:lattice_parameters}
\end{figure*}

Figures~\ref{fig:lattice_parameters}(a) and (b) show the lattice parameters as a function of
pressure. In Fig.~\ref{fig:lattice_parameters}(a), the $a$ parameter has been scaled by
$1/\!\sqrt{8}$, as the orthorhombic lattice parameters would become $b=c=a/\!\sqrt{8}$ in a
hypothetical NaCl-type structure \cite{WC79}. At low pressures, SnSe is most compressible
along its $c$-axis, but this direction becomes significantly less compressible at around the
critical pressure. In contrast, the $b$-direction exhibits initially a relatively low
compressibility, and it becomes more compressible just above the critical pressure. The
difference between the $b$ and $c$ parameter diminishes under pressure up to the critical
pressure and then increases again for higher pressures. The axial ratios shown in
Fig.~\ref{fig:lattice_parameters}(c) provide an alternative illustration of these
observations. All three axial ratios have extrema near the critical pressure of 10.5~GPa. The
lattice of SnSe approaches a tetragonal metric at the critical pressure, but remains
orthorhombic at all pressures.

\subsection{Symmetrisation of the Crystal Structure}

\begin{figure}
     \centering
	  \includegraphics[scale=1.0]{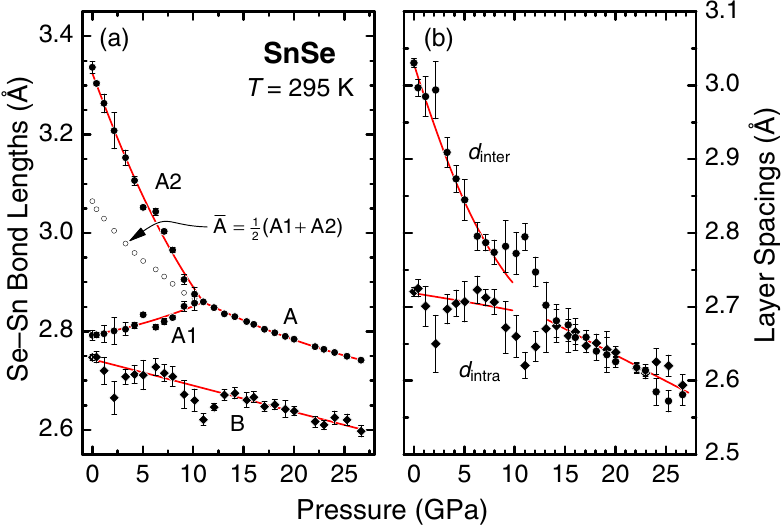}\\
     \caption{(a) Selected Se--Sn bond lengths and (b) the spacings between atomic sheets
              within and between the double layers in SnSe as a function of pressure. The
              symbols denoting the different distances are explained in
              Fig.~\ref{fig:crystal_structure}. Solid lines are guides to the eye.}
     \label{fig:distances}
\end{figure}

The pressure dependences of three characteristic \mbox{Se--Sn} interatomic distances within
the double layers are shown in Figure~\ref{fig:distances}(a). With increasing pressure, the
Se--Sn bond A1 and contact A2, which are oriented approximately parallel to the layers
(Fig.~\ref{fig:crystal_structure}), equalise in length and become equivalent bonds in the
$\beta{'}$ phase above 10.5~GPa. Interestingly, the A1 bond lengthens with increasing pressure
up to the transition, which can be attributed to the increasing influence of the A2 contact.
Also noteworthy is the smooth transition from the average A1/A2 distance, $\bar{\mathrm{A}}$,
in the low-pressure phase to bond length A in the high pressure phase. The short Se--Sn bond
B, perpendicular to the layers, remains the shortest bond at all pressures, and decreases in
length at approximately the same rate as bond A above the transition pressure.

The symmetrisation of the structure manifests itself also in the variation of the spacings of
the atomic sheets within and between the double layers, $d_{\mathrm{intra}}$ and
$d_{\mathrm{inter}}$, as shown in Figure~\ref{fig:distances}(b). These two layer spacings
become indistinguishable above 10~GPa. The exact transition pressure cannot be determined from
these data because of the relatively large scatter of the data points around 10~GPa, which can
be traced to the extracted $x$ atomic coordinates (see Fig.~\ref{fig:atom_coords}), and we
consider these fluctuations to be an artefact. Overall, the symmetrisation of the crystal
structure is predominantly driven by the changes in atomic positions, with only secondary
contributions from the variations of the lattice parameters with pressure.

\subsection{Relation to the High-Temperature Phase $\beta$-SnSe}

When referring to previous work on SnSe and related systems, one should note that different
settings have been used for the orthorhombic space groups, in particular the combination of
space groups $Pbnm$ and $Cmcm$ for the structures at ambient conditions and high temperatures,
respectively. The transformation from $Pnma$ to $Pbnm$ (and also from $Bbmm$ to $Cmcm$)
changes the axes as $a{'},b{'},c{'}\, [Pbnm] = c,a,b\, [Pnma]$. We will continue to use the setting
of $Pnma/Bbmm$ and convert the results of previous studies to this setting where needed.

The pressure-induced structural changes in SnSe are similar to, and yet different from those
at high temperature and normal pressure \cite{WC79,SW81}. With increasing temperature at zero
pressure, the Se and Sn atoms move continuously to a higher-symmetry ($Bbmm$) position and the
difference between the $b$ and $c$ lattice parameters decreases just as observed here as a
function of pressure. However, above the critical temperature of 807~K, the $b$ and $c$
parameters become indistinguishable and the lattice pseudo-tetragonal, whereas the
high-pressure form described here remains orthorhombic at all pressures. Even though the
high-temperature phase, $\beta$-SnSe, and the high-pressure form are equivalent in terms of
their crystallographic symmetry, they differ in terms of their lattice metric and may thus be
considered distinct phases. Therefore, we denote the ambient-temperature, high-pressure
modification as \mbox{$\beta{'}$-SnSe}.

\subsection{$P$--$V$ Equation of State}

\begin{figure}[b]
     \centering
	  \includegraphics[scale=1.0]{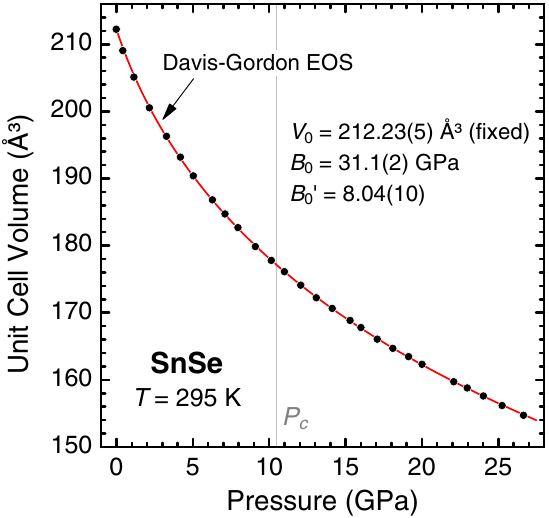}\\
     \caption{Pressure-volume relation of SnSe. The symbols show the measured volumes, and
              the solid line represents the best-fitting Davis-Gordon equation of state as
              described in the text.}
     \label{fig:EOS}
\end{figure}

The variation of the unit cell volume of SnSe with pressure is shown in Fig.~\ref{fig:EOS}. It
does not exhibit any discontinuities or other anomalies. Initial attempts to describe the
pressure-volume data with the Birch or Vinet equations of state yielded unsatisfactory
agreement between the fitted curves and the measured data, and the thus determined bulk
modulus and its derivative at zero pressure depended significantly on the range of pressures
that was used in the least-squares fit. The cause of these problems is the rather large change
in compressibility at low pressures. Therefore, we employed the less commonly used
Davis-Gordon equation of state \cite{DG67}, which was first applied to the pressure--volume
relation of mercury. It is a simple quadratic expansion of pressure $P$ in terms of the
relative change in density, $\Delta \rho/\rho_0 = (\rho-\rho_0)/\rho_0$. When expressed in
terms of volume $V$,
\begin{equation}\label{eq:DG_P(V)}
   P(V) = B_0(\Delta V/V) + \case{1}{2} B_0 (B_0^\prime -1 )(\Delta V/V)^2 \;,
\end{equation}
where $\Delta V /V = (V_0-V)/V$, and $B_0$ and $B_0^\prime$ are the bulk modulus and its
pressure derivative at zero pressure, respectively. Equation (\ref{eq:DG_P(V)}) can be
rearranged to yield
\begin{equation}
   V(P) = V_0 \,\frac{B_0^\prime -1}{B_0^\prime - 2 + \sqrt{\rule{0mm}{1.6ex} 1 + (B_0^\prime -1)\,2P/B_0}} \;.
\end{equation}
The Davis-Gordon equation of state describes the pressure--volume data for SnSe better than
the Vinet and Birch relations, and a fit with the $V(P)$ form gave $B_0=31.1(2)$~GPa and
$B_0^\prime = 8.04(10)$ with $V_0$ fixed at the measured zero-pressure volume, $V_0 =
212.23(5)$~{\AA}{\ensuremath{^3}}.

We find SnSe to be significantly more compressible than reported by Chattopadhyay \etal\
\cite{CWSP84}, who found $B_0 = 50.3(5)$~GPa (and $B_0^\prime = 6.3$). This discrepancy
persists even if we use, like Chattopadhyay \etal, the Birch equation of state to describe our
data, which gives $B_0 \approx 35$~GPa (and $B_0^\prime \approx 5.7$). On the other hand, our
measured bulk modulus is in fair agreement with Alptekin{'}s as well as Makinistian and
Albanesi{'}s computational results for \aSnSe\ \cite{Alp11,MA11}.

\subsection{Comparison with Related Systems}

The orthorhombic GeS and TlI structure types are well-known as possible intermediate
structures in reconstructive transformations between the NaCl and the CsCl structure types in
group-IV chalcogenides \cite{TKED03,OFS86,DKJP10}, but the understanding of the systematics of
the transitions in this group is still fragmentary. A systematic review of the vast literature
on the high-pressure behaviour of group-IV chalcogenides is beyond the scope of this work, but
we would like to highlight a few interesting cases. A pressure-induced transition from GeS- to
TlI-type phases is a recurring theme in recent computational work on GeS \cite{Dur05}, GeSe
\cite{GGJM14}, SnS \cite{AD10}, and SnSe itself \cite{Alp11}. X-ray diffraction studies on GeS
\cite{CWSP84} and GeSe \cite{CWSP84,OSFM97} did not yield evidence of such a transition, but
the quality of the experimental data would probably not have permitted detection of such a
relatively subtle transition.

In x-ray diffraction experiments on SnS, Ehm \mbox{\etal}\ observed an abrupt phase transition
between 16 and 18~GPa \cite{EKDK04}. A monoclinic lattice and space group $P2_1/c$ were
tentatively assigned to the high-pressure phase, and on this basis a volume jump of $\Delta
V/V \approx -9\%$ was inferred to occur at the transition. Even though the quality of the
x-ray diffraction data was insufficient to fully determine the crystal structure of the
high-pressure phase, these findings are incompatible with the computational results
\cite{AD10}. It is worth noting that in the computational investigations the crystal symmetry
often appears to have been restricted to space group $Pnma$ (which includes the
higher-symmetry $Cmcm$ as a special case) so that no conclusions can be drawn with regards to
transitions to other structure types beyond GeS and TlI. Examples of such alternative
structure types are the orthorhombic structures reported for PbTe \cite{RKSR05} and  SnTe
\cite{ZLMC13} at pressures of 6--16~GPa and 4--18~GPa, respectively.

\section{Conclusions}

We have determined the structural evolution of SnSe under hydrostatic pressure up to 27 GPa
and observed a continuous transition from the GeS to the TlI structure type with a critical
pressure of 10.5(3)~GPa. There are no indications of any discontinuities in the structural
parameters as a function of pressure. The high-pressure phase $\beta{'}$-SnSe is closely related
to the high-temperature phase at ambient pressure, $\beta$-SnSe, but whereas the former
remains orthorhombic at all pressures, the latter has a tetragonal metric of the crystal
lattice. A transition of this kind, with an increase of the crystal symmetry from space group
$Pnma$ to $Cmcm$, has been a recurring outcome of computational studies on group-IV
chalcogenides. To the best of our knowledge, the present study provides the first experimental
evidence for such a transition.

The similarity between the high-pressure and the high-temperature phases leads us to expect
that the stability field of the high-temperature, higher-symmetry modification will extend to
lower temperatures as pressure is applied. This suggests that under the application of
pressure or compressive strain, SnSe may exhibit very good thermoelectric properties at
temperatures lower than the $\sim$800~K that are required for bulk SnSe at normal pressure. It
appears therefore worthwhile to explore whether using strained films of SnSe may permit to
lower the operating temperature of SnSe-based devices while maintaining excellent
thermoelectric performance, which would open a route to overcoming the problem of sublimation
and hence limited thermal stability of SnSe at high temperatures of $\sim$800~K.

\section*{Acknowledgments}
We thank M. Hanfland (ESRF) and R. Howie (Edinburgh) for their assistance. Facilities were
made available by the European Synchrotron Radiation Facility, and part of this work was
supported by the Leverhulme Trust (RPG-2012-576).

\end{document}